\documentclass[sigconf, authordraft, review=false ,authorversion,nonacm]{acmart}
\usepackage{multirow}
\usepackage{orcidlink}
\AtBeginDocument{%
  }

\begin{document}

\title{Real-Time Assistive Navigation for the Visually Impaired: A Scalable Approach for Indoor and Outdoor Mobility}

\author{Dabbrata Das \orcidlink{0009-0008-0049-2048}}
\authornote{These authors contributed equally to this work.}
\email{dasdabbrata@gmail.com}
\affiliation{%
  \institution{Khulna University of Engineering \& Technology}
  \city{Khulna}
  \country{Bangladesh}}
  
\author{Argho Deb Das \orcidlink{0009-0000-5756-2483}}
\authornotemark[1]
\email{atdeb727@gmail.com}
\affiliation{%
  \institution{Northern University Bangladesh}
  \city{Dhaka}
  \country{Bangladesh}}

\author{Farhan Sadaf \orcidlink{0009-0007-7550-7972}}
\authornotemark[1]
\authornote{Corresponding author.}
\email{farhansadaf@cse.kuet.ac.bd}
\affiliation{%
  \institution{Khulna University of Engineering \& Technology}
  \city{Khulna}
  \country{Bangladesh}}

\author{Azhar Uddin \orcidlink{0009-0002-3692-7492}}
\email{amicableamit105112@gmail.com}
\affiliation{%
  \institution{Khulna University of Engineering \& Technology}
  \city{Khulna}
  \country{Bangladesh}}

  \author{Tirtho Mondal \orcidlink{0009-0003-3314-9100}}
\email{tirthomondal.2001@gmail.com}
\affiliation{%
  \institution{Khulna University of Engineering \& Technology}
  \city{Khulna}
  \country{Bangladesh}}

\renewcommand{\shortauthors}{Das et al.}

\begin{abstract}
Navigating unfamiliar environments remains one of the most persistent and critical challenges for people who are blind or have limited vision (BLV). Existing assistive tools often rely on online services or APIs, making them costly, internet-dependent, and less reliable in real-time use. To address these limitations, we propose PathFinder, a novel mapless mobile phone-based navigation system that operates fully offline. Our method processes monocular depth images and applies an efficient pathfinding algorithm to identify the longest, clearest obstacle-free route, ensuring optimal navigation with low computational cost. Comparative evaluations show that PathFinder reduces mean absolute error (MAE), speeds decision-making, and achieves real-time responsiveness indoors and outdoors. A usability study with 15 BLV participants confirmed its practicality, where 73\% learned to operate it in under a minute, and 80\% praised its accuracy, responsiveness, and convenience. Despite challenges in complex indoor layouts and low light, PathFinder offers a low-cost, scalable, reliable alternative.
\end{abstract}

\ccsdesc[500]{Human-centered computing~Accessibility technologies}  
\ccsdesc[300]{Computing methodologies~Computer vision problems}  
\ccsdesc[100]{Information systems~Navigation-based services}

\keywords{Blind and low vision (BLV) people, Indoor and outdoor navigation, Assistive technology, Large Language Models (LLMs), Pathfinder Algorithm, Computer Vision}

\maketitle


\section{Introduction}
Navigating unfamiliar environments is a significant challenge for blind and low-vision (BLV) individuals, both indoors and outdoors. While traditional mobility aids such as white canes and guide dogs provide essential support, these have limitations in detecting obstacles and identifying clear paths in complex environments. To enhance mobility, BLV individuals rely on assistive technologies such as GPS-based navigation systems, AI-powered smart glasses \cite{intro_smartGlass}, and sonar-equipped smart canes \cite{intro_sonarSmartCane}. These solutions provide audio or haptic feedback for obstacle detection and route guidance. However, despite these advancements, existing technologies \cite{cscw_nav} still face limitations in real-time navigation, precise obstacle avoidance, and environmental awareness, particularly in dynamic and unfamiliar scenarios.

Many GPS-based assistive technologies, including \textit{Be My Eyes} \cite{intro_bemyeyes}, \textit{Seeing AI} \cite{intro_seeingAI}, and \textit{Aira} \cite{intro_aira}, offer voice guidance \cite{voice_guide} and real-time assistance. However, they often struggle with providing precise localization, real-time obstacle avoidance, and effective pathfinding in unstructured environments. Additionally, AI-powered solutions, such as computer vision-based smart glasses \cite{intro_deepLearningSmartGlass}, require high computational power, making them costly and less accessible for many users. Given these challenges, there is a need for a more efficient, low-cost, and user-friendly navigation solution that leverages image processing and depth estimation techniques to assist BLV individuals effectively.

In this paper, we examine various navigation techniques, including those that use Vision Language Models (VLMs), and then present a computer vision-based algorithm to aid in BLV navigation. Our proposed system incorporates monocular depth estimation to generate depth maps, which, when processed using LLMs, improve obstacle detection and free path identification. Finally, we propose a novel image-processing approach that applies an innovative pathfinding algorithm on depth images to identify the longest free path, ensuring optimal route selection from mobile phone camera. This computationally efficient technique enhances real-time navigation by providing accurate obstacle-free paths while maintaining low processing costs.

To evaluate the effectiveness of our system, we aim to answer the following research questions:
\begin{itemize}
    \item \textbf{RQ1:} How useful is our proposed navigation scheme for BLV people?
    \item \textbf{RQ2:} How effective are AI-powered solutions for real-time navigation, and how does our method address their shortcomings?
    \item \textbf{RQ3:} How well does the implemented system address BLV individuals' needs, and what are their impressions of its effectiveness?
\end{itemize}

To answer RQ1, we conducted a qualitative analysis of our mobile application  (Section~\ref{sec:usability_study} ) with BLV individuals to explore why they would rather choose our proposed navigation system over existing methods, such as white canes, guide dogs, or standard GPS-based applications. The findings revealed recurring challenges in current practices, including difficulties in anticipating environmental layouts, reliably detecting obstacles, and maintaining orientation in dynamic settings. Participants emphasized the need for a tool that not only enhances safety but also supports independence without requiring specialized hardware or extensive training. These insights guided the design of our system, which leverages a single-camera smartphone to provide an accessible, user-friendly, and real-time navigation solution.

To answer RQ2, we performed a quantitative comparison (Section~\ref{sec:results}) between our method and existing AI-powered navigation solutions. While AI-based systems can assist navigation, results show that they do not yield significant improvements in navigation performance relative to their computational and financial costs. Many of these solutions depend on cloud-based APIs, which incur recurring expenses, introduce latency, and limit scalability for widespread adoption. In contrast, our system is designed to operate fully offline, eliminating the need for internet access, thereby ensuring scalability across diverse contexts and making it more affordable and reliable for BLV users globally. By integrating depth maps with a novel pathfinding algorithm, our approach improves accuracy and reduces processing time, offering a more practical alternative to resource-intensive AI solutions.

To answer RQ3, we have conducted user studies and surveys (Section~\ref{sec:usability_study}) to assess our system’s effectiveness, usability, and long-term adoption potential. Early results indicate that BLV individuals value the system’s independence from constant connectivity and its ability to address real-world navigation challenges, such as dynamic obstacles and unfamiliar environments. In line with broader developments in assistive technology and policy, we also account for concerns around privacy, data security, and adaptability. Incorporating these considerations strengthens user trust and ensures the system’s relevance for long-term use.

While our system demonstrates notable strengths in accessibility, efficiency, and scalability, we recognize potential limitations related to adaptability in highly dynamic environments and the need for continuous user feedback. 

\section{Related Works}
Developing successful navigation systems for blind and low-vision (BLV) people has been a major focus of assistive technology research. Various methodologies have been investigated, ranging from GPS-based solutions and depth-sensing techniques to AI-powered systems and infrastructure-free navigation tools. While many of these technologies strive to increase mobility, obstacle detection, and spatial awareness, they frequently confront hurdles in real-time processing, environmental adaptability, and user accessibility. This section examines existing research on BLV navigation systems, depth estimate methods, and AI-powered assistive technologies, emphasizing their contributions, limits, and implications for the development of our proposed solution.

\subsection{Navigation Systems for BLV Individuals}

Initiatives into developing effective navigation aid for BLV individuals have evolved over time, from GPS-based solutions to infrastructure-free mobile applications and robotic aids, each targeting a unique facet of mobility and environmental awareness.

Early GPS-based systems, like Microsoft Soundscape \cite{rw_Microsoft_Soundscape}, intended to improve spatial awareness by providing 3D audio cues.  Liu et al. \cite{rw1} discovered that spatial audio callouts increased usability and retention, but also revealed difficulties in precise localization and real-time obstacle avoidance.  These disadvantages have prompted academics to investigate alternative ways to standard GPS.

One approach has been to utilize existing infrastructure and sensing capabilities. Research conducted by Jain et al. \cite{rw9} showed that street cameras can give real-time audio feedback, helping BLV pedestrians avoid barriers and cross roadways safely.  While promising, this strategy is primarily reliant on public monitoring infrastructure, posing questions about scalability and adoption. Wayfinding and Backtracking apps developed by Tsai et al.'s \cite{rw12}, use smartphone inertial sensors to facilitate pathfinding and retracing paths, are examples of infrastructure-free solutions for indoor environments.  These solutions eliminate the requirement for bluetooth beacons or any fixed infrastructure, but they are prone to localization drift, particularly in huge or featureless areas.

Other researchers have investigated depth sensing and embodied systems.  See et al. \cite{rw13} created a smartphone app that utilizes built-in depth cameras to identify obstacles and recognize objects, providing real-time environmental awareness without specialist hardware.  Kuribayashi et al. \cite{rw5} developed PathFinder, a suitcase-shaped robotic assistant that recognizes junctions and signage to direct users indoors.  While these approaches boost accessibility, they also present trade-offs, such as performance unpredictability in diverse situations and greater cognitive burden due to route remembering.  Similarly, Ren et al. \cite{rw6} built RouteNav, a hybrid system integrating GPS localization with inertial dead reckoning, which increased accuracy in structured areas like transportation hubs but still struggled with drift and signal discrepancies.

CrossingGuard \cite{rw16_CrossingGuard} has shown that providing more intersection-level information (e.g., traffic signals, road width, and intersection geometry) greatly boosted BLV pedestrians' confidence when crossing streets.  The study also proved the viability of scaling such information via crowdsourcing.  More contemporary systems, such as StreetNav \cite{rw15_StreetNav}, reuse existing street cameras to give exact outdoor navigation.  StreetNav reduced swerving, led participants closer to their destinations, and addressed privacy concerns by relying on existing infrastructure.  Together, these works show a path from early intersection-focused systems to modern infrastructure-based precision, emphasizing both the benefits and remaining constraints of scale, environmental variability, and occlusion in outdoor navigation.

\subsection{AI-Powered Scene Description and Accessibility Tools}

AI-powered scene description systems have been a key factor in improving BLV users' awareness of their surroundings.  These computer vision and NLP-based solutions seek to promote independence by delivering real-time descriptions of objects, people, and situations. However, trust, accuracy, and usability remain critical consideration factors.

Penuela et al. \cite{rw2} examined how BLV users engage with such applications, showing that while participants relied on AI-generated descriptions in daily life, errors and inconsistencies created tension between trust and perceived utility. The study emphasized the importance of improving accuracy and supporting customization to foster long-term adoption. Likewise, See et al. \cite{rw13} demonstrated how depth cameras can provide real-time obstacle awareness through auditory and haptic feedback without requiring additional hardware, though performance varied across contexts and object categories.

More recent research has expanded AI accessibility to areas such as data visualization.  Seo et al. \cite{rw8} developed maidrAI, a multimodal system that combines text, audio, and haptics to assist BLV users in understanding complex charts.  Participants praised clear, structured outputs and personalization elements, while issues such as AI hallucinations and prompt complexity restricted usage.  ChartA11y \cite{rw7} allows users to engage with scatter plots, bar charts, and line charts through multimodal smartphone interfaces in a related effort. User tests demonstrated the promise of touch-based exploration mixed with aural feedback, but gesture complexity and small-scale evaluations revealed areas for improvement.

This collection of works demonstrates both the promise and the problems of AI-powered accessibility aids.  While these technologies can increase independence, these solutions face recurring challenges with trust, reliability, and user control that must be solved before widespread real-world adoption.

\subsection{User Engagement and Broader Assistive Technologies}
Designing effective navigation and accessibility technologies needs more than just technical accuracy; it also depends on how people embrace, interact with, and use these systems over time.  A increasing body of research highlights the value of user-centered design, personalization, and consideration of the larger social and educational contexts in which blind and low-vision (BLV) individuals use assistive technology.

Liu et al. \cite{rw1} investigated user behavior with Microsoft Soundscape to find parameters influencing prolonged engagement. Their findings revealed that early experiences, such as the first week of use, were highly predictive of long-term adoption.  Features like spatial audio callouts and previews were crucial in keeping consumers interested.  Tsai et al. \cite{rw12} highlighted the importance of flexible, infrastructure-free navigation tools.  Their Wayfinding and Backtracking programs enabled hands-free mobility without the need for extra hardware; however, limitations such as drift limited their general usefulness.

Aside from turn-by-turn navigation, researchers have proposed that assistance technologies should promote exploration and spatial learning.  According to Jain et al.'s study \cite{rw17}, BLV users frequently seek higher-level spatial knowledge, such as the overall shape and structure of places, to construct cognitive maps, rather than depending exclusively on directions.  The study also found that preferences for information varied across individuals based on blindness onset, sociability, and mobility aid use, emphasizing the importance of customisation.  It emphasized the collaborative element of exploration, in which BLV users routinely interact with trainers, peers, and spectators.  This viewpoint reframes navigation systems as components of a larger ecology of interconnectedness, rather than simply tools for independence.

Research has also demonstrated how expanded accessibility tools influence learning and inclusiveness.  Shaheen \cite{rw10} examined BLV students' problems in K-12 schools and discovered that many obtained assistive technology (AT) literacy outside of school due to lack of formal education assistance.  Complex interfaces and unlabeled buttons were particularly detrimental to STEM student participation.  Hoogsteen et al. \cite{rw11} found that intersections, traffic signals, and crosswalks were highly desired for urban navigation. Information needs differed between early-blind and late-blind users.  These findings highlight that content selection, user training, and long-term adoption are just as important as technical innovation.

Taken together, this body of studies demonstrates that good assistive technology cannot be evaluated solely on technological parameters.  Early user experiences, support for exploration and cognitive mapping, sensitivity to social collaboration, and integration into larger educational and urban contexts are all critical for long-term adoption.

\subsection{Gaps in Existing Research}

Despite substantial advances in assistive navigation and AI-powered accessibility technology, major gaps remain, limiting their real-world impact for blind and low-vision (BLV) people.  

First, several approaches encounter computational and hardware limitations.  Depth estimation and object identification algorithms frequently need significant computing resources, rendering them unsuitable for reliable internet-free deployment. Moreover, these systems frequently rely on specialized hardware such as depth cameras or LiDAR, which raises costs and limits accessibility \cite{rw13}.

Second, many systems struggle with environmental flexibility.  Navigation solutions that rely on pre-mapped patterns or static environments fail in dynamic urban settings where barriers are continuously shifting.  Infrastructure-based solutions, such as those employing public cameras \cite{rw9}, show potential but confront issues with scalability and large-scale dependability.  Similarly, inertial-based indoor navigation systems are susceptible to time drift, diminishing long-term accuracy \cite{rw12}.

Third, concerns about trust and information quality persist.  BLV users frequently become frustrated when systems deliver excessive or misleading information.  AI hallucinations, which provide false or overly detailed responses, might exacerbate these challenges and diminish trust in key situations \cite{rw2,rw8}. 

Fourth, personalization and user flexibility are frequently undeveloped.  Many technologies use a one-size-fits-all concept, ignoring variations in blindness onset, mobility tactics, and individual preferences.  Multimodal interfaces (e.g., touch, audio, haptics) have improved accessibility for data visualization \cite{rw7}, but they still fail to meet various user needs.  Studies reveal that early-blind and late-blind users generally require different types of environmental information, yet most systems fail to account for these distinctions \cite{rw11}.
 
Finally, the larger challenges of scalability and inclusion remain unaddressed.  Educational environments, for example, continue to underserve BLV learners, with students frequently learning assistive technology literacy outside formal teaching \cite{rw10}.  This absence of universally adaptable and inclusive solutions contributes to ongoing marginalization and emphasizes the importance of designs that are adaptable across contexts, from schools to busy urban streets.

These limitations suggest that future research should look beyond improving technological accuracy and instead address concerns of trust, personalization, social context, and equity.  These gaps are both technological and sociotechnical, emphasizing the importance of efficiency, adaptability, usability, and inclusive system design.  This entails designing navigation and accessibility systems not only as technological artifacts, but also as components of a larger interconnected ecosystem in which usability, collaboration, and inclusivity are equally as vital as algorithmic performance.

\section{Methodology}
This section describes the navigation approaches explored in this study, including different pathfinding schemes, AI-powered scene description, and our proposed image-processing-based technique for BLV individuals' real-time navigation.
\subsection{Navigation Schemes}
To guide BLV individuals effectively, we consider three navigation schemes that define movement direction and obstacle-free pathfinding:
\subsubsection{Left/Forward/Right (DOF 3)}
This approach provides three degrees of freedom (DOF), allowing users to navigate in left, forward, or right directions based on obstacle-free regions. It simplifies decision-making but may lack precision in complex environments.
\subsubsection{In Degree (DOF 180)}
A more granular approach, where the free path is determined across a 180-degree field of view, offering finer directional control for more precise navigation.
\subsubsection{Clock Hour Hand (DOF 13)}
Inspired by the clock-face representation, this method divides the navigation space into 13 distinct directions, mimicking how blind individuals often conceptualize orientation (e.g., "move toward 12 o'clock" for straight ahead). This approach balances accuracy and user intuitiveness.

\begin{figure}[h]
  \centering
  \includegraphics[width=\linewidth]{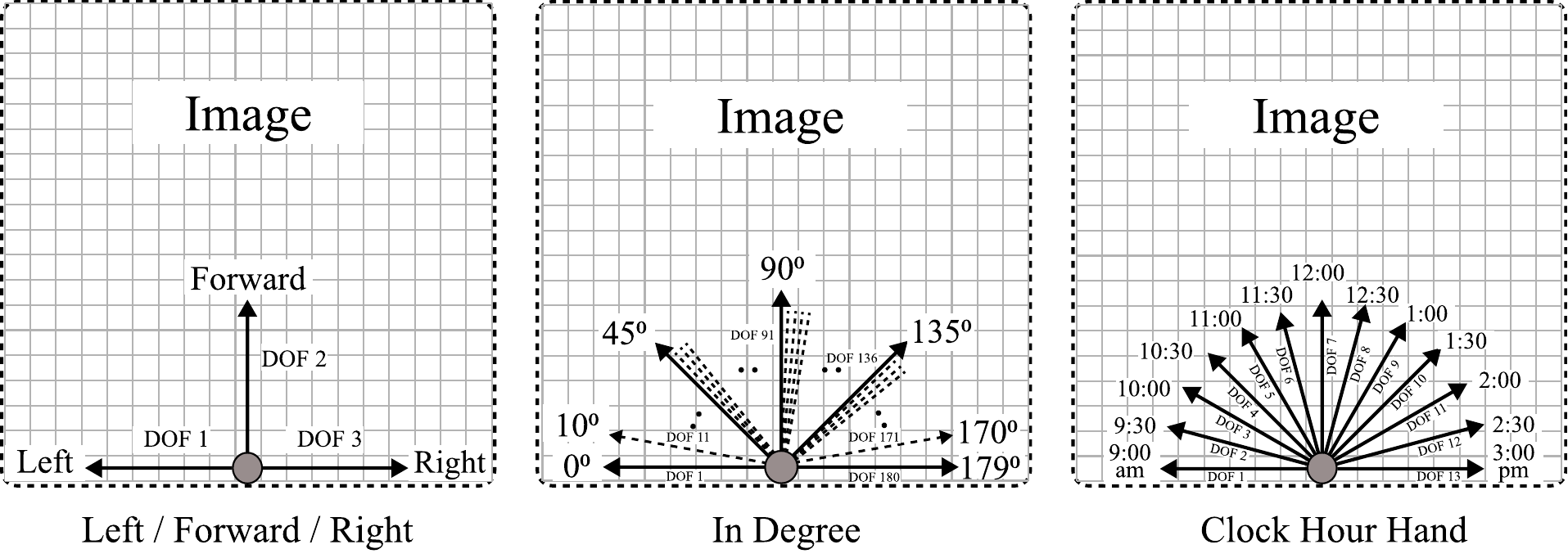}
  \caption{Visualization of three navigation schemes to define navigation direction.}
  \label{fig:navigation_scheme}  
  \Description{A woman and a girl in white dresses sit in an open car.}
\end{figure}

\begin{figure}[h]
  \centering
  \includegraphics[width=\linewidth]{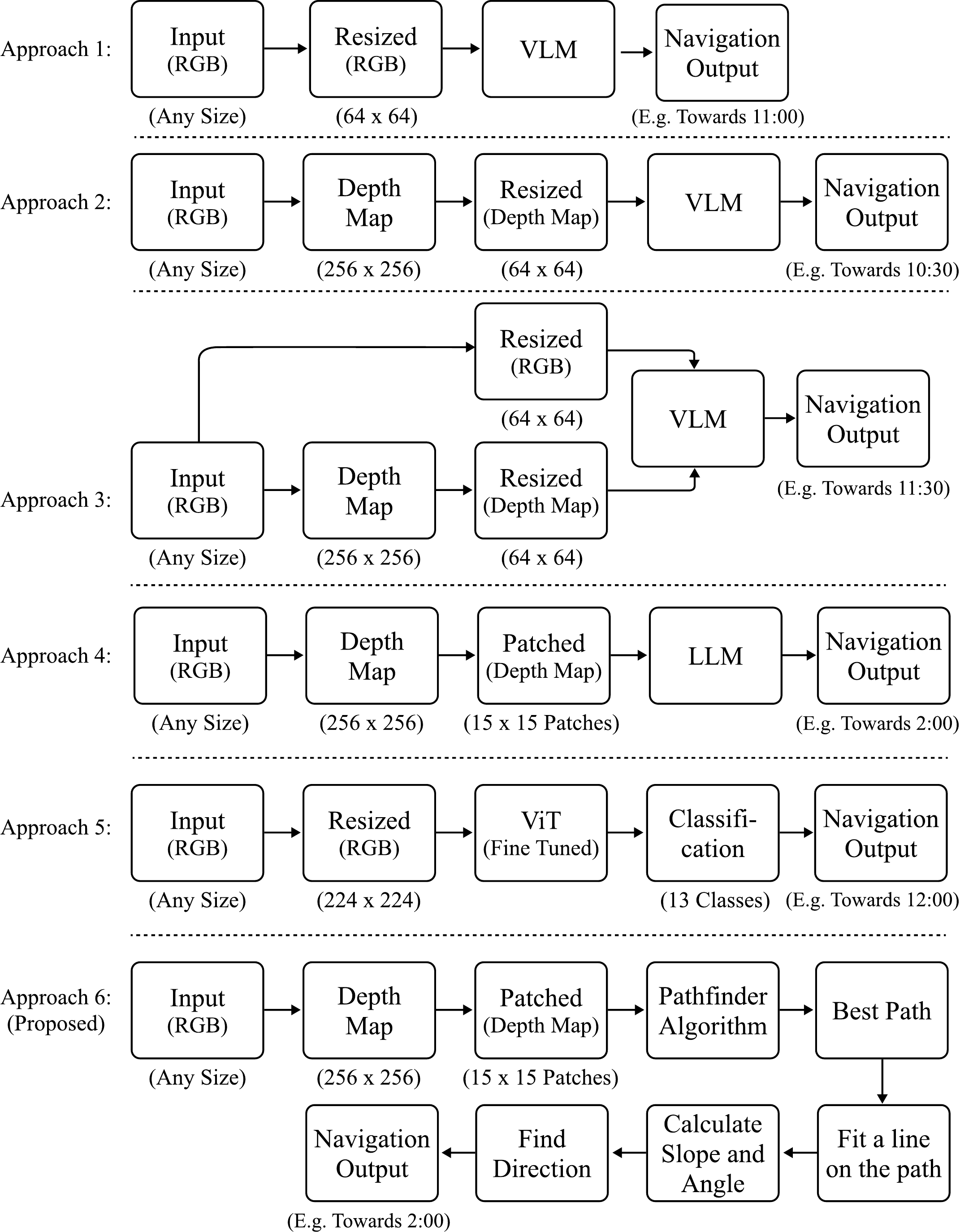}
  \caption{Block diagram of all our approaches for identifying the best path for BLV navigation.}
  \label{fig:proposed_approach_visualization}  
  \Description{A woman and a girl in white dresses sit in an open car.}
\end{figure}

In this study, explore six different navigation approaches, five of which leverage AI models for obstacle avoidance and pathfinding. In contrast, the proposed Approach 6 introduces a computationally efficient image-processing algorithm that employs a novel pathfinding technique to determine the longest free path. These approaches differ in terms processing methods, and model complexity, with the goal of balancing accuracy, efficiency, and real-time usability.
\subsection{AI-Assisted Navigation Approaches}
\subsubsection{\textbf{Approach 1: 2D RGB Image to Vision Language Models (VLMs)}}
In this approach, we use 2D RGB images with a resolution of 64×64 pixels to simplify processing, reduce computational complexity, and optimize inference time. The images are then passed to Vision Language Models (VLMs), which analyze the scene and infer the best path for navigation based on visual features. While this method benefits from VLMs' reasoning capabilities, its reliance on RGB data limits depth perception and precise obstacle avoidance.
\subsubsection{\textbf{Approach 2: Depth Images to VLMs}}
64×64 depth images are used instead of RGB images to enhance depth perception. Depth information allows VLMs to understand spatial relationships more effectively, improving obstacle detection and navigation decisions. VLMs still require significant computational power for real-time processing, making this approach less efficient for low-resource settings.
\subsubsection{\textbf{Approach 3: Combined RGB and Depth Images to VLMs}}
This approach combines 64×64 RGB and depth images, providing VLMs with visual and depth information for pathfinding. The model gains a richer understanding of the environment by fusing these two modalities, leading to improved accuracy. Consequently, the increased input complexity results in higher computational costs, making real-time implementation more challenging.
\subsubsection{\textbf{Approach 4: Patch-Based Depth Representation to LLMs}}
Instead of using full-resolution depth images, this method processes depth data in a 15×15 grid of patch values, significantly reducing input size while retaining essential spatial details. The processed depth patches are fed into LLMs, which analyze the segmented depth information for navigation. Although this method lowers computational requirements, the reduction in resolution may impact accuracy in complex environments.

\subsubsection{\textbf{Approach 5: Classification-Based Pathfinding using Vision Transformers (ViT)}}
\begin{figure}[h]
    \centering
    \includegraphics[width=1.0\linewidth]{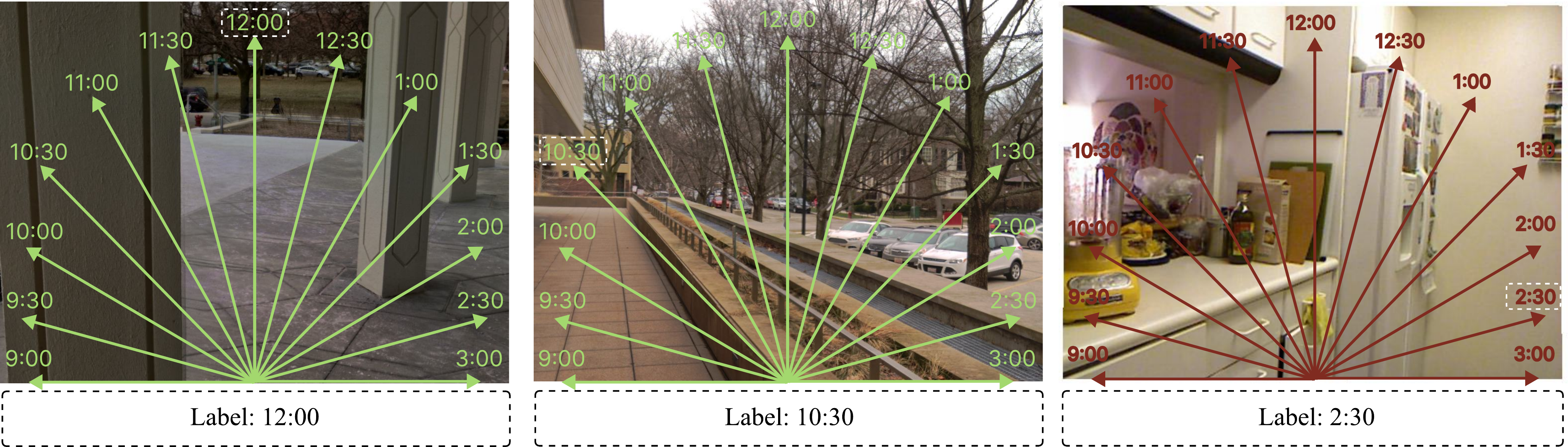}
    \caption{Direction annotation using a clock-hand overlay, where the nearest free space to the virtual hour hand is labeled as the ground-truth navigation direction.}
    \label{fig:direction_annotation}
\end{figure}

In this approach, pathfinding is formulated as a classification problem, where the system predicts the best navigation direction from an input image. To achieve this, we developed a custom dataset labeled with clock-hand directions, representing the free space available in each scene. A visual annotation tool was designed that overlays a virtual clock hand spanning from the 9 o’clock to the 3 o’clock positions on the image. Annotators selected the direction of free space that aligned most closely with the hour-hand, and this was recorded as the ground-truth navigation direction (Figure~\ref{fig:direction_annotation}).  

\begin{figure}[h]
    \centering
    \includegraphics[width=0.8\linewidth]{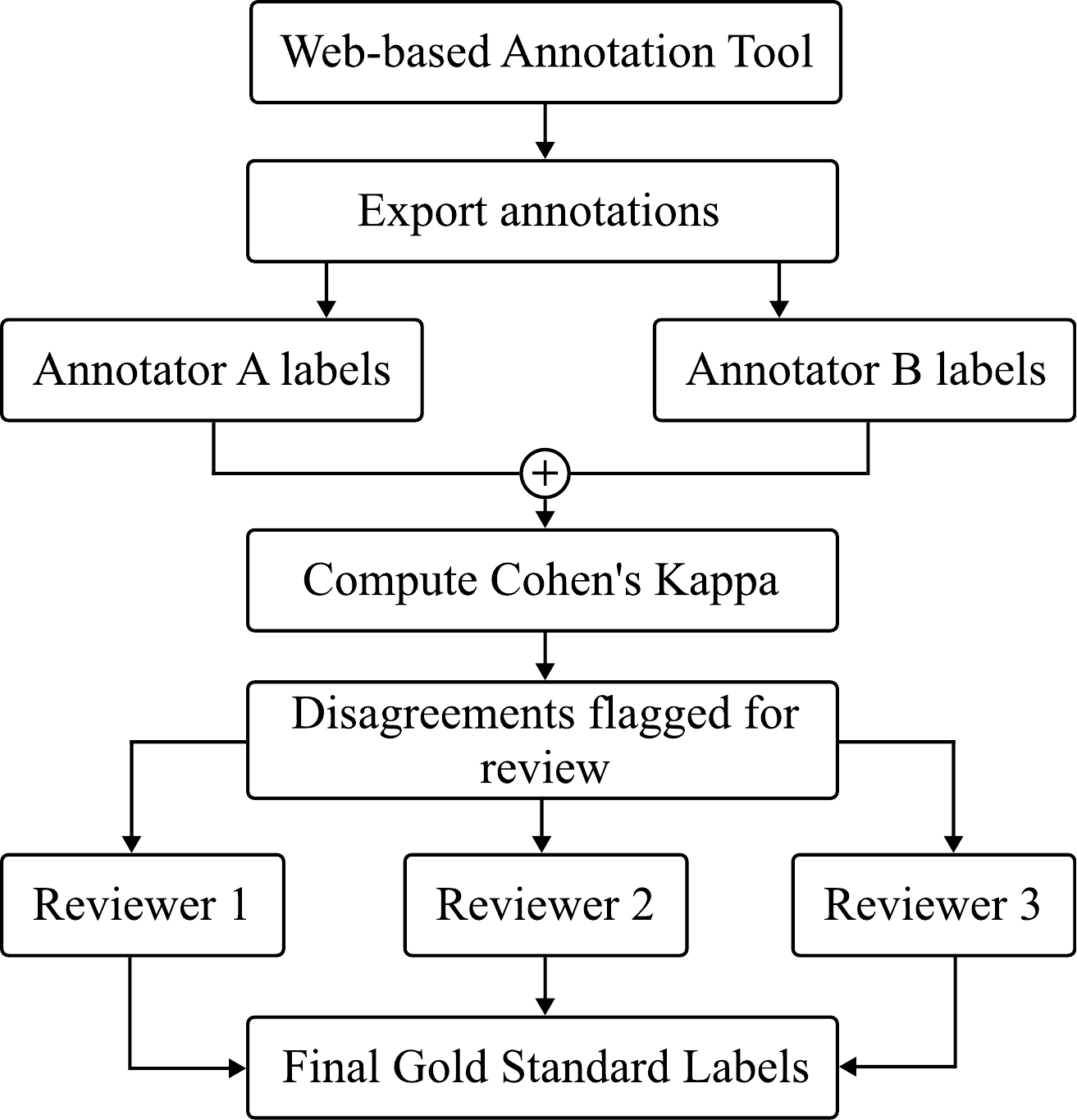}
    \caption{End-to-end annotation workflow used to generate the dataset labels.}
    \label{fig:annotation_workflow}
\end{figure}

Two annotators manually labeled each image using a web-based interface, and their annotations were cross-verified using Cohen’s kappa agreement procedure~\cite{cohens_kappa} by three more reviwers. They further inspected the dataset and validated the final labels. The resulting dataset comprised 270 training images and 30 testing images. Figure~\ref{fig:annotation_workflow} depicts the end-to-end annotation process used to generate the dataset labels. 

We then employed a supervised learning strategy, fine-tuning the Vision Transformer (ViT-Base) model on this annotated dataset. The model was trained to learn the mapping between extracted image features and the labeled navigation directions. While ViTs are effective at capturing long-range dependencies within images, their requirement for significant computational resources and large-scale data makes them less practical for real-time deployment in assistive navigation contexts. Nonetheless, this experiment demonstrates the feasibility of formulating pathfinding as a classification task and highlights the trade-offs between model complexity and deployment efficiency.

\subsubsection{\textbf{Approach 6: Novel Pathfinding from Patched Depth Images (Proposed)}}
While AI-assisted solutions have demonstrated strong performance in complex perception tasks, they often suffer from significant limitations in real-world applications. These deep learning-based approaches typically demand substantial computational resources, rely on high-end hardware, and consume considerable energy—making them inefficient and unsuitable for real-time or on-device use, particularly in low-power or embedded systems. Additionally, they often require large annotated datasets for training and may struggle to generalize well across diverse, unseen environments.

To address these challenges, we propose a computationally efficient image processing algorithm for pathfinding. Our method leverages depth images obtained from monocular depth estimation and utilizes a 15×15 patch-based depth representation. Instead of relying on AI models, we implement a pathfinding algorithm to traverse the depth image and identify the longest obstacle-free path, ensuring a clear and safe route for navigation.

\begin{figure}[h]
  \centering
  \includegraphics[width=\linewidth]{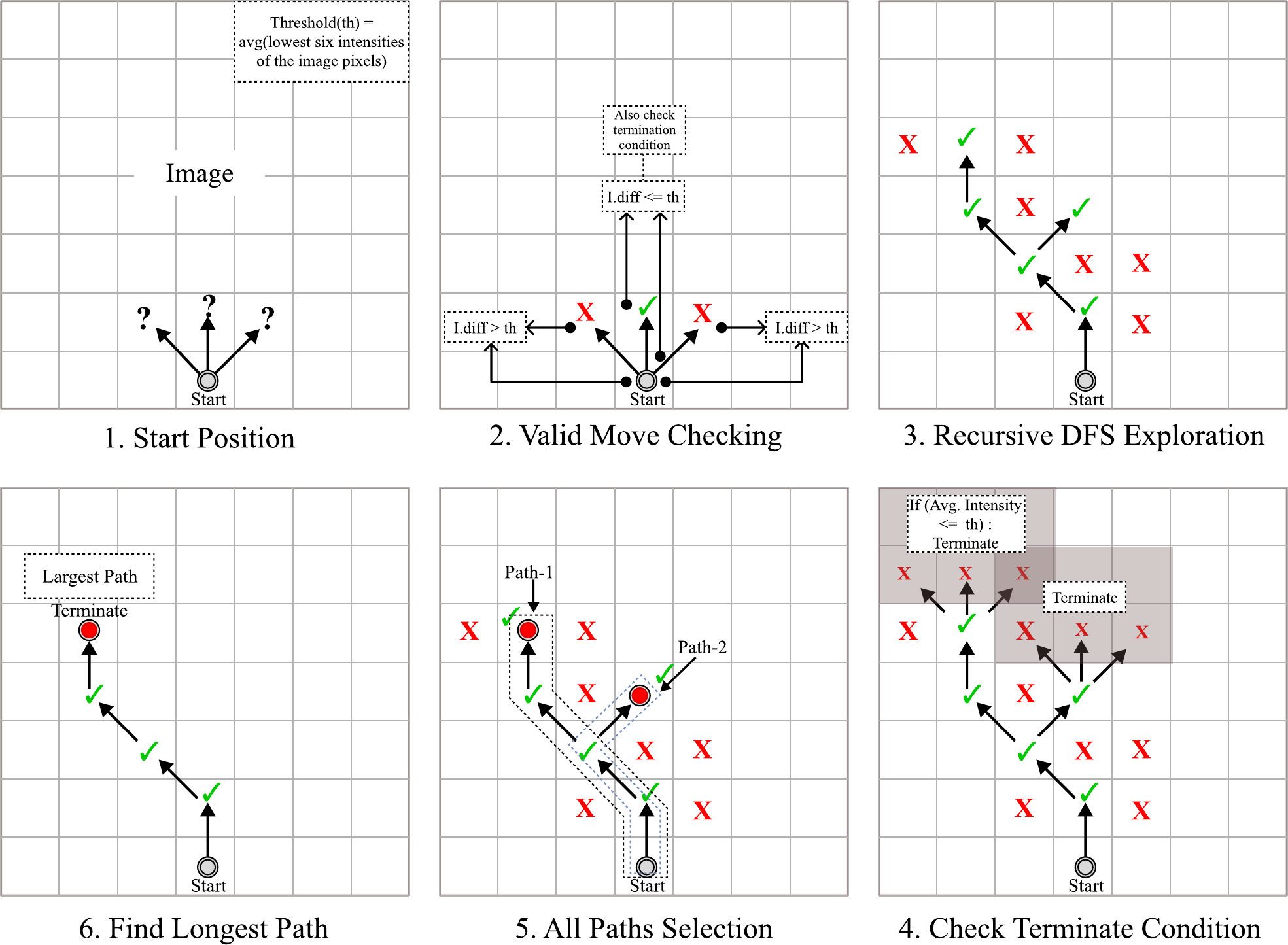}
  \caption{Process for optimal pathfinding calculation illustrating path exploration and optimal route selection.}
  \label{fig:algo_diagram_fig}
\end{figure}

The algorithm is designed based on two key assumptions:
    \begin{itemize}
        \item \textbf{First}, the depth scene generally transitions from light to dark, meaning the algorithm will prioritize paths that follow a decreasing intensity trend.
        \item \textbf{Second}, the search starts from the lower bottom middle of the image, as this corresponds to the natural position of a phone held in front of a person.
    \end{itemize}
The search progresses upwards, exploring valid paths while adhering to these constraints.
The diagram (Figure \ref{fig:algo_diagram_fig}) provides a detailed representation of the algorithm.\\

Optimal Pathfinding Algorithm:

\begin{tabbing}
\hspace{0.3cm}\= \hspace{0.5cm}\= \hspace{0.7cm}\= \hspace{0.9cm}\= \kill
\textbf{FUNCTION} PathFinder(matrix, start\_row, start\_col, threshold) \\
\> \textbf{DEFINE} recursive search procedure to explore paths: \\
\>\> ADD current position to the path \\
\>\> COMPUTE avg\_intensity of six upper pixels: \\
\>\> \hspace{0.4cm} (up-left, up-up-left, up, up-up, up-right, up-up-right) \\
\>\> \textbf{IF} avg\_intensity $<$ threshold \textbf{OR} no valid moves \textbf{THEN} \\
\>\>\> ADD path to results and \textbf{RETURN} \\
\>\> \textbf{IF} top rows reached \textbf{OR} intensity diff. $>$ threshold \textbf{THEN} \\
\>\>\> ADD path to results and \textbf{RETURN} \\
\>\> \textbf{FOR EACH} valid neighbor (up, up-left, up-right) \textbf{DO} \\
\>\>\> \textbf{IF} neighbor value $\leq$ current value \textbf{THEN} \\
\>\>\>\> RECURSE search with updated path \\
\>\> \textbf{IF} no further moves \textbf{THEN} \\
\>\>\> ADD path to results \\
\> CALL search from (start\_row, start\_col) \\
\> \textbf{RETURN} the longest and most straight-forward path
\end{tabbing}

\subsection{Mobile Application Prototype}
We created a smartphone application to test our system's usability in real-world scenarios.  The application takes input from the device camera and uses the suggested depth estimation and pathfinding pipeline locally to ensure offline functioning.  The interface was built with few buttons and supports multimodal feedback, including speech via the native text-to-speech engine, directional vibration signals, and optional visual overlays illustrated in Figure \ref{fig:feedback_interface}.

To evaluate system responsiveness, the application was tested in common indoor and outdoor environments.

\section{Results and Discussion}
\label{sec:results}

\begin{table*}[h]
    \centering
    \caption{Performance Comparison for Indoor Scenes}
    \label{tab:indoor_scenes}
    \resizebox{\textwidth}{!}{%
        \begin{tabular}{|l|l|ccc|ccc|}
            \hline
            \multirow{2}{*}{Approaches} & \multirow{2}{*}{Model} 
              & \multicolumn{3}{c|}{Tested on 300 Images} 
              & \multicolumn{3}{c|}{Tested on 30 Images} \\
            \cline{3-8}
            & & MAE (degree) $\downarrow$ 
                & Accuracy (\%) $\uparrow$ 
                & Avg.\ Response Time (s) $\downarrow$ 
              & MAE (degree) $\downarrow$ 
                & Accuracy (\%) $\uparrow$ 
                & Avg.\ Response Time (s) $\downarrow$ \\
            \hline
            Approach 1 (Real(64x64) $\rightarrow$ VLM) 
              & gemini-1.5-pro & 49.98 & 15.67 & 4.306 
                               & 47.28 & 15.44 & 4.324 \\
            & BLIP-2       & 56.02 & 10.08 & 3.288 & 55.50 & 12.00 & 3.260 \\
            & LLaVA-1.5    & 58.78 &  9.44 & 3.040 & 58.26 &  9.98 & 3.110 \\
            \hline
            Approach 2 (Depth(64x64) $\rightarrow$ VLM)
              & gemini-1.5-pro & 46.04 & 15.45 & 4.776 
                               & 44.66 & 16.92 & 4.825 \\
            & BLIP-2       & 48.00 & 13.98 & 3.562 & 47.78 & 14.67 & 3.616 \\
            & LLaVA-1.5    & 49.32 & 12.67 & 3.041 & 50.02 & 10.78 & 3.120 \\
            \hline
            Approach 3 (Real+Depth $\rightarrow$ VLM)
              & gemini-1.5-pro & 42.78 & 16.04 & 5.118 
                               & 42.00 & 16.67 & 5.161 \\
            & BLIP-2       & 47.24 & 14.32 & 3.340 & 46.50 & 14.75 & 3.360 \\
            & LLaVA-1.5    & 48.91 & 11.08 & 3.016 & 48.77 & 10.62 & 2.997 \\
            \hline
            Approach 4 (Depth(15x15(patch)) $\rightarrow$ LLM)
              & Phi-3.5-mini-instruct & 47.90 & 18.33 & 3.257 
                                     & 40.00 & 32.38 & 3.257 \\
            & Phi-3-mini-4k-instruct & 54.35 & 14.00 & 3.094 
                                      & 51.00 & 20.00 & 3.094 \\
            & gemini-1.5-pro         & \textbf{39.05} & 20.67 & 5.750 
                                      & 37.50 & 30.00 & 5.750 \\
            \hline
            Approach 5 (Classification $\rightarrow$ ViT) Train270\_Test30
              & ViT-Base & --    & --    & --    
                         & 52.50 & 13.29 & \textbf{0.087} \\
            \hline
            \textbf{Approach 6 (Depth(15x15(patch)) $\rightarrow$ Pathfinder Algo)(Proposed)} 
              & Pathfinder Algorithm 
              & 39.81 & \textbf{32.04} & \textbf{0.377} 
              & \textbf{25.78} & \textbf{37.67} & 0.399 \\
            \hline
        \end{tabular}%
    }
\end{table*}

\begin{table*}[h]
    \centering
    \caption{Performance Comparison for Outdoor Scenes}
    \label{tab:outdoor_scenes}
    \resizebox{\textwidth}{!}{%
        \begin{tabular}{|l|l|ccc|ccc|}
            \hline
            \multirow{2}{*}{Approaches} & \multirow{2}{*}{Model} & \multicolumn{3}{c|}{Tested on 300 Images} & \multicolumn{3}{c|}{Tested on 30 Images} \\
            \cline{3-8}
            & & MAE (degree) $\downarrow$ & Accuracy (\%) $\uparrow$ & Avg. Response Time (s) $\downarrow$ & MAE (degree) $\downarrow$ & Accuracy (\%) $\uparrow$ & Avg. Response Time (s) $\downarrow$ \\
            \hline
            Approach 1 (Real(64x64) $\rightarrow$ VLM) & gemini-1.5-pro & 39.94 & 16.04 & 4.45 & 40.50 & 15.00 & 4.4497 \\
            & BLIP-2 & 40.67 & 14.72 & 3.65 & 42.00 & 14.00 & 3.65 \\
            & LLaVA-1.5 & 47.00 & 10.12 & 3.288 & 48.50 & 8.66 & 3.288 \\
            \hline
            Approach 2 (Depth(64x64) $\rightarrow$ VLM) & gemini-1.5-pro & 39.18 & 15.5 & 5.1912 & 40.00 & 13.33 & 5.1912 \\
            & BLIP-2 & 41.24 & 11.67 & 3.69 & 41.67 & 11.14 & 3.69 \\
            & LLaVA-1.5 & 47.58 & 10.38 & 3.32 & 47.62 & 9.95 & 3.32 \\
            \hline
            Approach 3 (Real(64x64) + Depth(64x64) $\rightarrow$ VLM) & gemini-1.5-pro & 40.50 & 14.66 & 5.109 & 41.74 & 13.04 & 5.109 \\
            & BLIP-2 & 42.39 & 14.00 & 3.557 & 44.00 & 12.78 & 3.557 \\
            & LLaVA-1.5 & 49.45 & 11.32 & 3.42 & 52.46 & 7.66 & 3.42 \\
            \hline
            Approach 4 (Depth(15x15(patch)) $\rightarrow$ LLM) & Phi-3.5-mini-instruct & 40.50 & 15.33 & 3.72 & 42.00 & 13.33 & 3.72 \\
            & Phi-3-mini-4k-instruct & 49.05 & 10.00 & 3.635 & 54.00 & 6.67 & 3.635 \\
            & gemini-1.5-pro & 40.13 & 16.05 & 5.615 & 41.00 & 16.67 & 5.615 \\
            \hline
            Approach 5 (Classification $\rightarrow$ ViT) (Train270\_Test30) & ViT-Base & -- & -- & -- & 41.38 & 13.79 & \textbf{0.091} \\
            \hline
            \textbf{Approach 6 (Depth(15x15(patch)) $\rightarrow$ Pathfinder Algo)(Proposed)} & Pathfinder Algorithm & \textbf{36.45} & \textbf{34.67} & \textbf{1.136} & \textbf{26.00} & \textbf{38.33} & 1.332 \\
            \hline
        \end{tabular}%
    }
\end{table*}

\subsection{Evaluation Metrics}

To evaluate the performance of the proposed directional prediction algorithm, we consider three key metrics: \textbf{Mean Absolute Error (MAE)}, \textbf{Accuracy}, and \textbf{Average Response Time}. We evaluated the performance with our annotated dataset (Section 3.2.5). The ground truth directions were manually annotated separately for 300 images from the NYU Depth V2 \cite{nyu} dataset (indoor scenes) and 300 images from the DIODE \cite{diode} dataset (outdoor scenes).
\subsubsection{\textbf{Mean Absolute Error (MAE)}}
MAE measures the average deviation between the predicted navigational direction and the ground truth, both expressed in clock-hour format. For example, if the ground truth direction is annotated as 11:00 (i.e., $330^\circ$), and the predicted direction is 10:00 (i.e., $300^\circ$), then the angular deviation is $|330^\circ - 300^\circ| = 30^\circ$, which corresponds to a 1-hour error.
Formally, MAE is calculated as:
\begin{equation}
\text{MAE} = \frac{1}{N} \sum_{i=1}^{N} \left| \theta_i^{\text{pred}} - \theta_i^{\text{gt}} \right|
\end{equation}
where \( N \) is the total number of images, \( \theta_i^{\text{pred}} \) is the predicted direction in degrees for the \( i^{\text{th}} \) image, and \( \theta_i^{\text{gt}} \) is the corresponding ground truth direction.

\subsubsection{\textbf{Accuracy}}
Accuracy is defined as the proportion of predictions that exactly match the ground truth in clock-hour format. That is, if both predicted and annotated directions are the same (e.g., both are 2:00), it is considered a correct prediction.
\begin{equation}
\text{Accuracy} = \frac{1}{N} \sum_{i=1}^{N} \delta(\theta_i^{\text{pred}}, \theta_i^{\text{gt}})
\end{equation}
where the indicator function \( \delta(a, b) \) is defined as:
\begin{equation}
\delta(a, b) =
\begin{cases}
1, & \text{if } a = b \\
0, & \text{otherwise}
\end{cases}
\end{equation}

\subsubsection{\textbf{Average Response Time}}
The average response time refers to the mean computational time (in milliseconds) taken by the algorithm to generate the predicted direction for each input image. This metric reflects the system’s suitability for real-time applications.\\

Among the metrics, \textbf{MAE} is the most crucial as it directly quantifies angular deviation from the ground truth and thus determines the reliability of navigation. While high \textbf{Accuracy} ensures consistency, a low \textbf{MAE} guarantees that even near-misses are within acceptable tolerance levels. \textbf{Response Time}, though secondary, is significant for real-time deployment, particularly in assistive navigation applications for visually impaired users.

\begin{figure*}[h]
  \centering
  \includegraphics[width=0.8\linewidth]{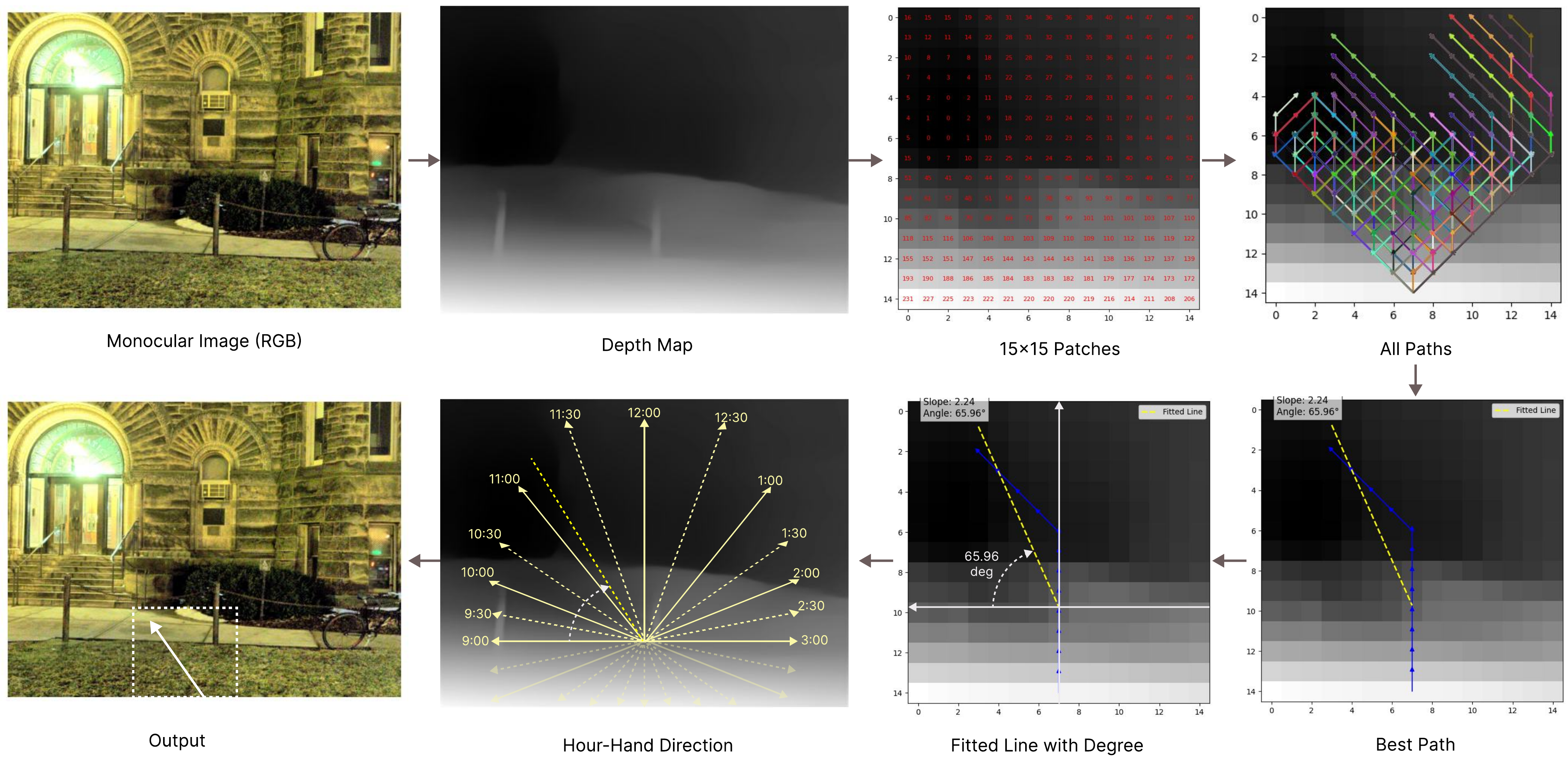}
  \caption{A step-by-step visualization of our proposed method for identifying the optimal path using the Pathfinder algorithm, applied to a sample input from the DIODE \cite{diode} dataset.}
  \label{fig:proposed_approach_visualization}  
  \Description{A woman and a girl in white dresses sit in an open car.}
\end{figure*}

\subsection{Indoor Scene Performance}

The performance comparison for indoor scenes, as shown in Table \ref{tab:indoor_scenes}, reveals key insights into the effectiveness and efficiency of the proposed and baseline approaches. Notably, the proposed \textbf{Approach 6 (Depth(15x15(patch)) $\rightarrow$ Pathfinder Algo)} demonstrates competitive performance across key metrics, balancing mean absolute error (MAE), accuracy, and average response time.

For the indoor scene, \textbf{Approach 6} achieves an MAE of \textbf{25.78 degrees} with an accuracy of \textbf{37.67\%} and a notably fast average response time of \textbf{0.399 seconds}. While the MAE value is slightly better than some LLM-based approaches (e.g., \textbf{Approach 4 (Depth(5x5(patch)) $\rightarrow$ LLM)} with \textbf{37.5 degrees} MAE using \textbf{gemini-1.5-pro}), the proposed method significantly outperforms others in terms of computational efficiency. The proposed approach achieves an average response time of \textbf{0.399 seconds}, which is significantly faster than VLM and LLM-based methods, where response times range between \textbf{3.094} and \textbf{5.75 seconds} for other approaches. This indicates that the Pathfinder Algorithm is highly optimized for rapid processing, making it a suitable choice for real-time navigation applications.

Although the classification-based \textbf{Approach 5 (ViT-Base)} exhibits the lowest response time of \textbf{0.087 seconds}, but its MAE of \textbf{52.5 degrees} and accuracy of \textbf{13.29\%} indicate a trade-off between speed and prediction quality. This suggests that while ViT-based approaches are computationally efficient, they may lack the precision required for fine-grained obstacle navigation tasks.

\subsection{Outdoor Scene Performance}

Table \ref{tab:outdoor_scenes} presents the performance comparison for outdoor scenes, where the proposed \textbf{Approach 6} again demonstrates remarkable efficiency and improved accuracy under specific conditions. For the set of 30 test images, \textbf{Approach 6} achieves an MAE of only \textbf{26.00 degrees} with an accuracy of \textbf{38.33\%}, outperforming all other approaches in both metrics. This represents a significant improvement over the best-performing VLM-based approach (\textbf{Approach 4 with gemini-1.5-pro}), which achieved an MAE of \textbf{41.00 degrees} and an accuracy of \textbf{16.67\%}.

The efficiency of the proposed approach is further highlighted by its response time of \textbf{1.332 seconds}, which, while slower than the ViT-based method (\textbf{0.091 seconds}), is still markedly faster than all VLM and LLM-based methods, which suggests that the Pathfinder algorithm strikes an effective balance between accuracy and computational efficiency, making it well-suited for real-world applications where rapid decision-making is critical.

Interestingly, the proposed \textbf{Approach 6} also performs consistently across indoor and outdoor environments, with only a slight increase in response time when transitioning to outdoor scenes. This consistency indicates the robustness of the pathfinding method in diverse environments, which is essential for navigation systems assisting blind and low-vision (BLV) individuals.

\begin{table*}[h]
    \centering
    \caption{Impact of Patch Size on Inference Time, MAE, and Accuracy for Indoor~\cite{nyu} and Outdoor~\cite{diode} Scenes}
    \label{tab:patch_size_full_comparison}
    \begin{tabular}{|c|c|c|c|c|c|c|}
        \hline
        \textbf{Patch Size} 
        & \multicolumn{3}{c|}{\textbf{Indoor Scenes}} 
        & \multicolumn{3}{c|}{\textbf{Outdoor Scenes}} \\
        \cline{2-7}
        & Avg. Response Time (s) & MAE (degree) $\downarrow$ & Accuracy (\%) $\uparrow$ 
        & Avg. Response Time (s) & MAE (degree) $\downarrow$ & Accuracy (\%) $\uparrow$ \\
        \hline
        480 $\times$ 480 & 0.267 & 52.80 & 25.12 & 0.727 & 50.95 & 28.81 \\
        240 $\times$ 240 & 0.294 & 46.50 & 28.84 & 0.831 & 44.80 & 30.42 \\
        120 $\times$ 120 & 0.317 & 44.24 & 30.50 & 0.922 & 41.20 & 32.40 \\
        60  $\times$ 60  & 0.340 & 42.95 & 31.21 & 0.967 & 38.37 & 32.52 \\
        30  $\times$ 30  & 0.362 & 40.20 & \textbf{34.50} & 1.012 & 36.54 & 34.21 \\
        15  $\times$ 15  & 0.377 & \textbf{39.81} & 32.04 & 1.136 & \textbf{36.45} & \textbf{34.67} \\
        5  $\times$ 5  & 0.391 & 39.84 & 32.25 & 1.424 & 36.48 & 33.92 \\
        \hline
    \end{tabular}
\end{table*}

Based on the results in Table~\ref{tab:patch_size_full_comparison}, we adopt a 15$\times$15 patch size for our depth-based pathfinding framework. Patch size selection is crucial since depth maps are used for object localization, and meaningful detection requires grouping pixels into sufficiently large regions. Very large patches (e.g., 480$\times$480) result in faster inference but lose granularity, missing smaller or irregularly shaped obstacles. On the other hand, very small patches increase detection granularity but substantially raise inference time.  

The 15$\times$15 patch configuration achieves the best balance across metrics: it yields the lowest MAE for indoor scenes (39.81$^{\circ}$) and the highest accuracy for outdoor scenes (34.67\%), while maintaining response times within practical limits (0.377s indoors and 1.136s outdoors). This trade-off ensures that the system captures diverse object sizes with sufficient accuracy without incurring excessive computational cost, making it well-suited for real-time BLV navigation applications.

\subsection{Application Feedback}

The system provides navigation guidance through three complementary feedback modes: \textbf{voice}, \textbf{haptic}, and \textbf{visual}. This multimodal design ensures that blind and low-vision (BLV) users can receive clear instructions in diverse contexts, whether indoors, outdoors, quiet, or noisy.

\subsubsection{\textbf{Voice Feedback}}
Voice instructions are generated using the device’s Text-to-Speech (TTS) engine~\cite{tts}. Two formats are supported: (i) short prompts such as \textit{``move forward''}, and (ii) clock-based directions such as \textit{``2 o’clock''}. These concise formats allow users to quickly interpret orientation and act accordingly. As shown in Fig.~\ref{fig:feedback_interface}, audio and clock-direction cues worked best indoors, where quiet and structured spaces allowed participants to follow guidance naturally.

\subsubsection{\textbf{Haptic Feedback}}
To ensure reliable navigation in noisy environments, the application employs haptic guidance through vibration signals. Distinct vibration patterns and types, corresponding to clock-based directions, are mapped as summarized in Appendix A. Each pattern is unique, making it easy for users to distinguish direction solely through tactile feedback. When participants were outdoors, vibration cues proved more effective in cutting through environmental noise and distractions, while still providing accurate directional assistance.

\subsubsection{\textbf{Visual Feedback}}
Although unnecessary, we implemented visual feedback. For users with partial vision, where simple visual cues on the phone display, such as arrows (Fig. \ref{fig:feedback_interface}) or clock-hand overlays. This redundant channel reduces uncertainty and supports residual sight usage, making navigation smoother and more intuitive. Visual feedback complements audio and haptic guidance, offering a flexible third layer of support in both indoor and outdoor scenarios.

The strength of the system lies in its multimodal integration. While audio and visual cues are effective in structured indoor spaces, haptic guidance ensures clarity and safety in complex or noisy outdoor environments. Together, the three feedback channels increase flexibility, reduce confusion, and empower BLV users to navigate safely, confidently, and independently.

\begin{figure}[h]
  \centering
  \includegraphics[width=0.9\linewidth]{Photos/Feedback.pdf}
  \caption{PathFinder feedback interface showcasing (a) voice feedback, (b) haptic feedback, and (c) visual feedback.}
  \label{fig:feedback_interface}
\end{figure}

\subsection{Comparative Analysis and Key Observations}

\textbf{Accuracy vs. Efficiency Trade-off.} While VLM and LLM-based approaches achieve competitive accuracy in some cases, their response times are significantly higher, making them less practical for real-time navigation. The proposed pathfinding method (Approach 6) provides a favorable trade-off, offering a substantial improvement in response time (approximately 3-10 times faster) while maintaining competitive accuracy (around 2-3 times better), without a severe loss in performance. This demonstrates the method's ability to balance precision with efficiency, making it more suitable for real-time applications.
    
\textbf{Environment-Specific Performance.} The proposed method exhibits superior performance in outdoor scenes, achieving the lowest MAE and highest accuracy among all approaches. In particular, Approach 6 achieves a 106-152\% improvement in accuracy and a 16-24\% reduction in MAE compared to other methods, suggesting that the algorithm is particularly well-suited for more complex and variable environments like outdoor settings.

\textbf{Model-Specific Observations.} Among LLM-based methods, \textbf{gemini-1.5-pro} consistently outperforms other models (Phi-3.5-mini-instruct and Phi-3-mini-4k-instruct) in both accuracy and MAE. However, the increased computational cost associated with these models limits their real-time applicability. In contrast, the proposed pathfinding method achieves a similar or better performance in terms of accuracy, with a significantly lower response time (roughly 3-10 times faster), making it more suitable for practical deployment.
    
\textbf{Scalability and Real-Time Suitability.} The rapid processing times of the proposed method indicate its potential for deployment in real-time assistive navigation systems. Specifically, Approach 6 offers a 3-10 times faster response time than other methods, which is critical for time-sensitive applications. Furthermore, its robustness across both indoor and outdoor scenes suggests scalability to diverse environments, ensuring that the method remains effective in varying real-world conditions.

To sum up, the pathfinding approach offers several key advantages that make it especially suitable for real-time navigation assistance for blind or visually impaired (BLV) individuals. Its lightweight pathfinding algorithm demands significantly less computational power than conventional AI models, allowing efficient operation on mobile devices without high-end hardware. By utilizing depth variations, the method accurately distinguishes walkable paths from obstacles, enhancing path detection reliability. Additionally, it ensures real-time responsiveness by providing rapid path updates, which support seamless interaction in dynamic environments. Furthermore, the system is both user-friendly and cost-effective, requiring only a single-camera smartphone and eliminating the need for additional sensors or specialized equipment.

Overall, the proposed \textbf{Approach 6} emerges as a promising solution for assistive navigation, offering a balanced trade-off between accuracy and computational efficiency. The method stands out, particularly in outdoor environments where rapid decision-making is crucial, and it consistently delivers superior results compared to the alternatives.

\section{Usability Study}
\label{sec:usability_study}
To assess the usability of PathFinder and the Topline System, we conducted a study with 15 blind or visually impaired (BLV) participants from diverse age groups, education levels, mobility aids, and geographical locations. The study sought to answer the following critical questions:

\begin{itemize}
    \item How well does the established system meet the demands of BLV individuals, and how do they rate its effectiveness?
    \item How do alternative navigation schemes perform in terms of usability and accuracy?
    \item What were the biggest issues that participants encountered while using the app?
    \item What improvements or feature requests did the participants make?
\end{itemize}

\subsection{Participant Demographics and Mobility Aids}
The study included 15 participants, aged 25 to 80 years, with varying causes of blindness, including birth conditions, accidents, diabetic retinopathy, glaucoma, strokes, cataracts, and macular degeneration. Educational backgrounds ranged from no formal education to university graduates, and participants were from both urban and rural areas across Bangladesh (Table \ref{tab:participant_info}).

Participants used a variety of mobility aids, with the white cane (7 participants) being the most common, followed by smart canes (3), guide dogs (2), GPS-based smart assistants (2), and haptic feedback devices (2). These mobility aids influenced how participants interacted with the tested navigation systems and their expectations for usability and feedback mechanisms.

\begin{table}
  \caption{Summary of Survey Participants}
  \label{tab:participant_info}
  \centering
  \begin{tabular}{p{2.7cm} p{5cm}}
    \toprule
    \textbf{Factor} & \textbf{Details} \\
    \midrule
    Total Participants & 15 \\
    Age Range & 25 - 80 years \\
    Gender Distribution & 10 Male, 5 Female \\
    Cause of Vision Loss & Birth (2), Accident (2), Glaucoma (1), Diabetic Retinopathy (2), Stroke (2), Cataracts (2), Macular Degeneration (1), Retinitis Pigmentosa (1), Vitamin A Deficiency (1), Age-Related Vision Loss (1) \\
    Mobility Aids & White Cane (6), Smart Cane (3), Haptic Feedback Devices (2), Guide Dog (2), GPS-Based Smart Assistants (2) \\
    Testing Environment & Indoor (7), Outdoor (8) \\
    Average Duration of App Use & 10.13 minutes \\
  \bottomrule
\end{tabular}
\end{table}

\subsection{Participant Engagement and Feedback Collection}
To ensure a diverse evaluation context, we implemented various engagement strategies during the usability study of our application. We engaged in conversations with hospital patients, shopkeepers, neighbors, and community members. This approach provided insights into real-world navigation challenges and user preferences across different social settings. All participants were informed of the purpose of the testing and provided their consent to participate, ensuring that they understood their role in the evaluation process and the significance of their feedback.

\subsection{Ease of Use and Response Time}

Participants rated the ease of use of Approach 6 on a qualitative scale (\textit{Good, Average, Bad}):
\begin{itemize}
    \item \textbf{9} participants rated the system as \textbf{"Good"}.
    \item \textbf{4} participants rated it as \textbf{"Average"}.
    \item \textbf{2} participants rated it as \textbf{"Bad"}.
\end{itemize}

The response time for participants to complete navigation tasks ranged from \textbf{2 to 6 minutes}, with an average of approximately \textbf{3.8 minutes}. Participants using \textit{haptic feedback devices} or \textit{GPS-based smart assistants} generally took longer, whereas those using \textit{white canes} and \textit{guide dogs} exhibited more varied response times.

\subsection{Participant Feedback on Indoor and Outdoor Navigation}
Participants evaluated their navigation experience in \textbf{indoor} (7 participants) and \textbf{outdoor} (8 participants) settings using a five-point Likert scale, where \textbf{1 = Very Difficult} and \textbf{5 = Very Easy}. The responses were visualized through boxplots in figure \ref{fig:usability_plot_navbox}.

\subsubsection{\textbf{Indoor Navigation}}
Participants generally found indoor navigation more challenging, citing difficulties in identifying obstacles, turns, and structural variations.

\subsubsection{\textbf{Outdoor Navigation}}
Outdoor navigation received higher Likert ratings, as participants benefited from open spaces, familiar routes, and natural audio cues. However, a few participants rated outdoor navigation low due to traffic noise, moving obstacles, and surface irregularities.

\begin{figure}[h]
  \centering
  \includegraphics[width=\linewidth]{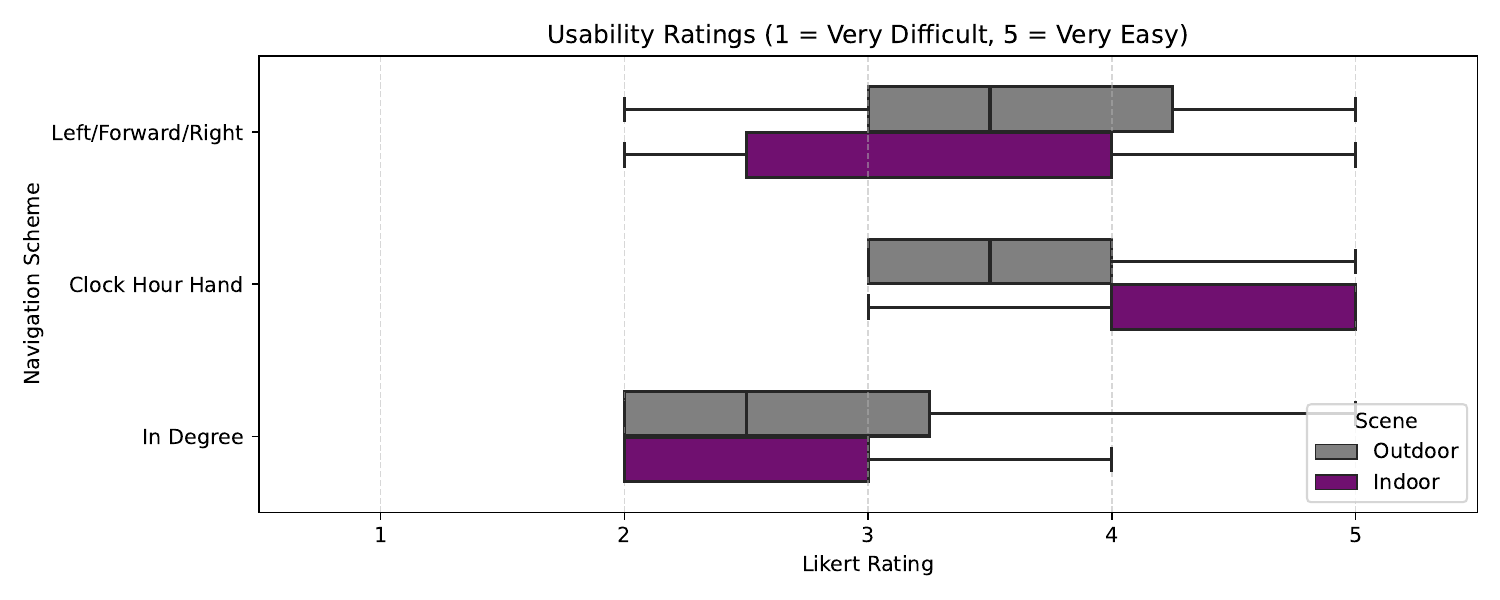}
  \caption{Usability ratings (1–5) for navigation schemes across indoor (purple) and outdoor (grey) scenes.}
  \label{fig:usability_plot_navbox}
\end{figure}

\subsection{User Preferences for Navigation Schemes}
To determine which navigation strategy felt most intuitive and practical for BLV users, participants ranked three distinct approaches: \textbf{Left/Forward/Right (DOF 3)}, \textbf{Clock Hour Hand (DOF 13)}, and \textbf{In Degree (DOF 180)}.  Participants rated these systems based on their ease of use, accuracy, and real-world application.  Ranking results are summarized as follows:
\begin{itemize}
    \item \textbf{Clock Hour Hand} scheme was selected as the first choice by 7 out of 15 participants, making it the most preferred option. Users praised its systematic approach to navigation, claiming that it closely corresponds with how they currently understand directions.
    \item \textbf{Left/Forward/Right} was ranked first by six participants, particularly those with minimal formal education or limited technological knowledge. They found it more intuitive since it resembled how a sighted guide would advise them. However, in crowded outdoor or cramped indoor environments, they did not prefer this method for long-term use.
    \item \textbf{In Degree} was the least popular option, with most participants claiming that precise angles were difficult to understand in real time. However, several individuals with technical skills or prior familiarity with digital navigation found it useful.
\end{itemize}

\begin{figure}[h]
  \centering
  \includegraphics[width=0.7\linewidth]{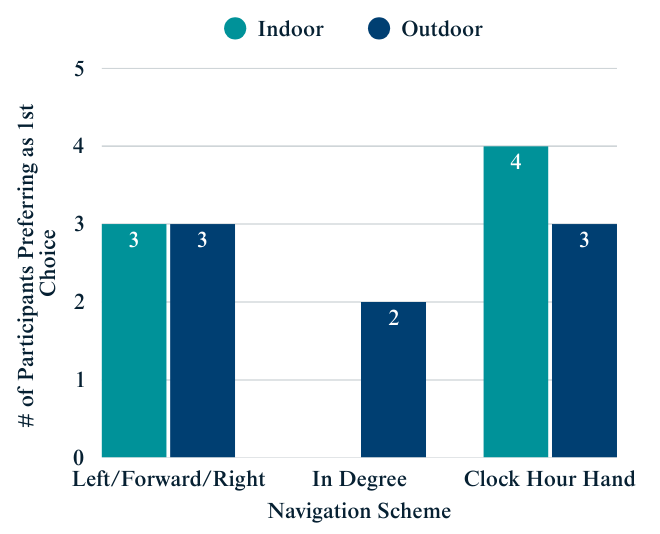}
  \caption{Bar chart showing the number of participants who selected each navigation scheme as their first priority, categorized by indoor and outdoor scenes.}
  \label{fig:usability_plot}
\end{figure}

\subsection{Method Ranking and Participant Preferences}
Participants ranked navigation methods based on usability and personal preference. Method 6 received the highest ranking across most participants, indicating its perceived effectiveness. Participants using haptic feedback devices and smart canes showed a preference for Method 5 and 6, while those using white canes had mixed preferences. Participants who rated ease of use as "Bad" had a more varied ranking of methods, suggesting difficulties in adapting to certain navigation approaches.
\subsubsection{\textbf{User Insights on Methods}}
Participants provided qualitative feedback on why they appreciated or disliked specific strategies.
Why Participants Preferred Method 6 (Pathfinder Algorithm):
\begin{itemize}
    \item "\textit{This way seems the most natural and accurate. I don't have to stop and think; I simply move.}"
    \item "\textit{It gives me clear paths quickly without making me confused.}"
    \item "\textit{Works well in crowded places where I need fast, precise decisions.}"
\end{itemize}

\begin{figure}[h]
  \centering
  \includegraphics[width=\linewidth]{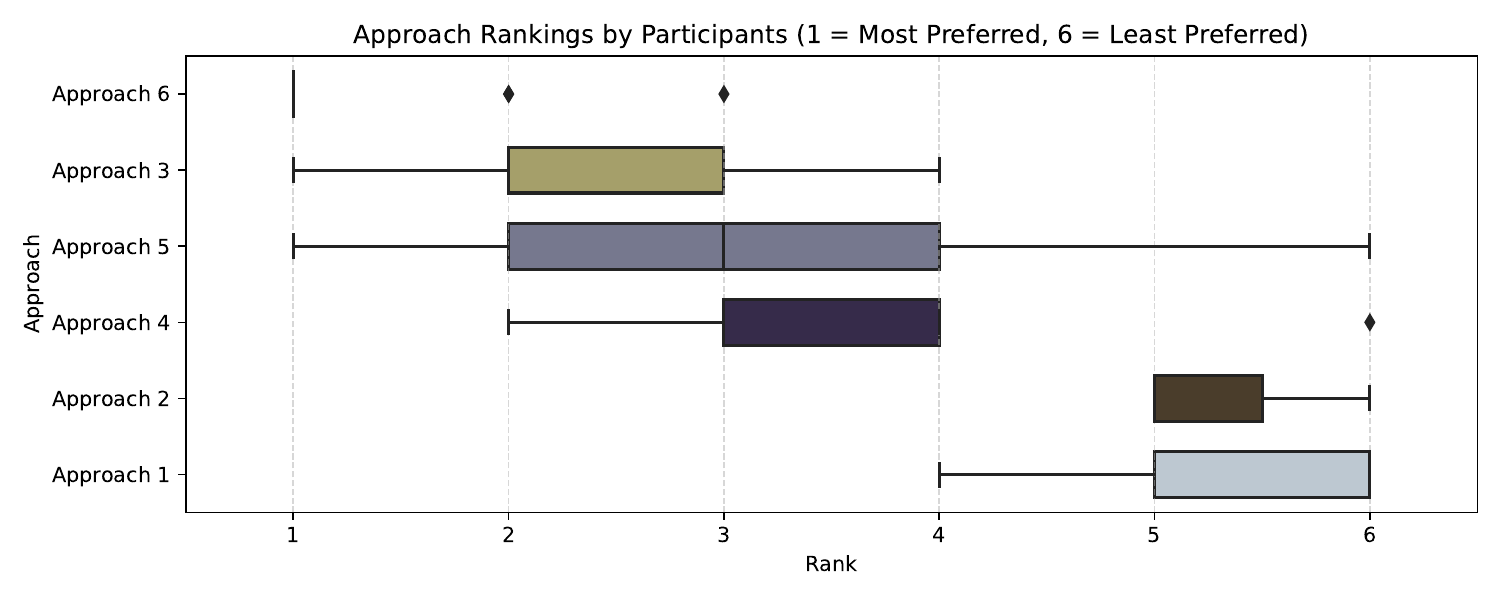}
  \caption{Rankings (1–6) of approaches across participants, with narrow boxplots showing rank distributions for each approach, where lower ranks indicate higher preference.}
  \label{fig:usability_plot_navbox}
\end{figure}

\subsection{Challenges with Other Approaches}
The table \ref{tab:participant_challenge_info} presents the challenges faced by all participants across various free pathfinding approaches.

\begin{table}[h!]

\caption{User challenges associated with various pathfinding methods.}
\label{tab:participant_challenge_info}
\centering
\renewcommand{\arraystretch}{1.2}
\begin{tabular}{p{2.1cm} p{5.5cm}}
\hline
\textbf{Approach} & \textbf{Challenges} \\
\hline

\textbf{Approach 1 \& 2} & "\textit{Too slow, sometimes the app took too long to respond. There was a delay between asking and getting a response, which made me hesitate. Good explanations, but the lag made it hard to use in real-time.}" \\

\textbf{Approach 3 \& 4} & "\textit{The app gave good directions, but sometimes it was unclear if I should turn or move forward. I liked the visual cues, but they were confusing in tight spaces. Works well when there's enough lighting, but struggles in dim areas.}" \\

\textbf{Approach 5 (ViT-Based Classification)} & "\textit{Fast but not always accurate, which is risky in unfamiliar areas. Quick to react, but occasionally misinterpreted obstacles. It missed a few important signs, which made me nervous while navigating.}" \\
\hline

\end{tabular}

\end{table}

\subsection{Learning the App \& Adaptability}
One of the most important components of usability is how rapidly BLV users learn and adapt to the application.  To evaluate this, participants provided input on Time Taken to Learn the App (How long it took them to feel comfortable using it), Training Needed (Whether they required assistance or self-learning was adequate), Feature Upgrade Requests (Any suggested enhancements for greater usability). Key Insights:
\begin{itemize}
    \item Table \ref{tab:learning_time} (Appendix B) shows that 66.67\% of participants learned the app in under a minute, indicating a straightforward design.

    \item \textbf{Users Who Did Not Need Training.} These participants, particularly those who were familiar with assistive technology, considered the app easy to use.

    \item \textbf{ Users Who Needed Training.} Some users (particularly elderly ones) preferred hands-on instruction before feeling comfortable using the program.

\end{itemize}

\begin{table}
  \caption{Learning Time Distribution of Participants}
  \label{tab:learning_time}
  \centering
  \begin{tabular}{p{2.5cm} p{2.8cm} p{2cm}}
    \toprule
    \textbf{Learning Time} & \textbf{Number of Participants} & \textbf{Percentage} \\
    \midrule
    $\leq$ 30 seconds & 1 & 6.67\% \\
    30-60 seconds & 9 & 60\% \\
    More than 1 min & 5 & 33.33\% \\
  \bottomrule
\end{tabular}
\end{table}

\subsection{Key Findings from the Usability Study}
Based on user input and review, several major insights arose about the app's usability and pathfinding methods for BLV persons.

\subsubsection{\textbf{What Worked Well}}
\begin{itemize}
    \item \textbf{Diverse navigation needs.} Mobility aids and environmental factors significantly influenced participant preferences.
    \item \textbf{Indoor navigation challenges.} Indoor environments presented more difficulties, reflected in lower Likert ratings and higher response times.
    \item \textbf{Outdoor navigation advantages.} Participants found outdoor navigation easier, though some faced challenges with environmental distractions.
    \item \textbf{Method 6 (Pathfinder Algorithm) was the clear favorite.} 80\% of participants (12 out of 15 from table \ref{tab:usability_study}) selecting it as their top priority due to various advantages such as good accuracy, speed, and ease of use.
\end{itemize}

\subsubsection{\textbf{Challenges Identified}}
\begin{itemize}
    \item \textbf{Some users had difficulty interpreting methods.} A few participants found certain navigation strategies, particularly in Degree (DOF 180), to be very complex.
    \item \textbf{33\% of participants required training}, with the majority being older or unfamiliar with technology.
    \item \textbf{Users requested clearer and more frequent voice instructions} to improve usability.
    \item \textbf{Response time variability.} While the app functioned well in most cases, some users reported slight delays in direction updates when obstacles changed rapidly.
    \item \textbf{Request for haptic feedback.} Users suggested incorporating haptic feedback, as voice output may not be practical in noisy or crowded places.
\end{itemize}

These findings underscore the importance of adaptive navigation solutions that accommodate different environments, mobility aids, and real-world challenges encountered by BLV individuals.

\begin{figure*}[h]
  \centering
  \includegraphics[width=0.8\linewidth]{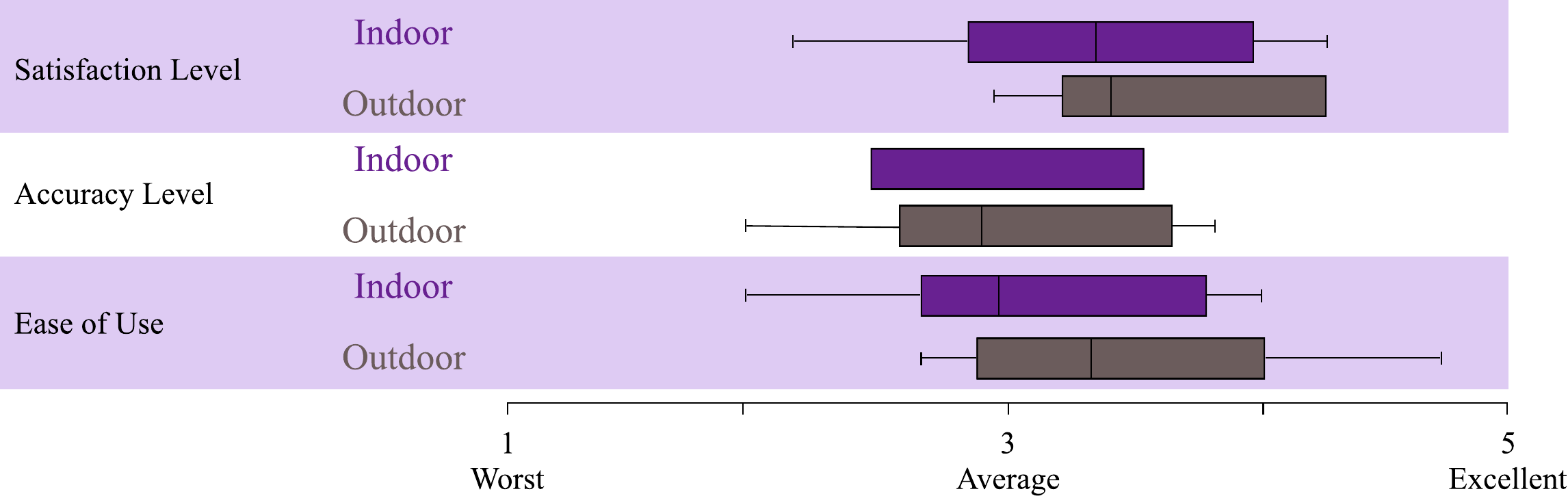}
  \caption{Feedback from a participatory study with 15 BLV patients, where responses on indoor (7 participants) and outdoor (8 participants) scenes were rated on a five-point Likert scale and visualized using boxplots to show percentile distributions.}
  \label{fig:usability_plot}
\end{figure*}

\section{Limitations}

One significant restriction is that our system does not direct people to their eventual destination, but rather offers a short, obstacle-free path in their immediate surroundings. While this enables safe navigation in local surroundings, incorporating a global path-planning mechanism may improve long-distance mobility and route optimization.

Another limitation is working in low-light circumstances, as the system is based on an image-based depth estimate. Depth accuracy diminishes in dark situations, making obstacle detection less accurate. Future enhancements could include the use of infrared or LiDAR-based sensors to increase performance in low-light conditions.

Although our technique doesn't struggle as much in dynamic situations, where moving objects like pedestrians and automobiles present unforeseen challenges. Our technology is intended for static scenarios, which limits its real-time adaptability. However, using haptic feedback instead of voice feedback might help resolve this issue, enabling more dynamic and responsive interaction. Future editions could investigate the use of real-time object tracking and motion prediction algorithms to deal with dynamic impediments more efficiently.

Furthermore, complex surfaces such as stairs and steep slopes cause difficulties for depth estimation, resulting in less precise path recommendations. To address this issue, the depth estimation model must be refined to better differentiate between navigable and non-navigable surfaces. Using terrain categorization algorithms may improve the system's capacity to manage such situations. Currently, there is no available dataset for this purpose. We have manually annotated a small dataset for testing, but a larger and more diverse dataset for fine-tuning could significantly improve the results of VLSs (Vision Language Systems).

Finally, participant recruitment for testing remains difficult, especially in Bangladesh, where many BLV users are unfamiliar with technology navigation aids. The limited availability to experienced users has an impact on the review process. Future development should concentrate on awareness campaigns and collaborations with local organizations to increase user adoption and collect more detailed feedback.

By addressing these limitations, we aim to improve our system's stability, usability, and real-world application, resulting in more independent mobility for BLV individuals.

\section{Conclusion}
This study presented a novel, computationally efficient navigation system designed to enhance mobility for Blind and Low Vision (BLV) individuals. By leveraging a depth-based pathfinding algorithm, we demonstrated a low-cost, real-time solution that accurately detects free paths in both indoor and outdoor environments. Comparative evaluations showed that our proposed system offers a favorable trade-off between accuracy and computational efficiency, making it a viable alternative to existing AI-powered navigation solutions that require high-end hardware.

Our usability study, conducted with 15 BLV participants, highlighted the system’s effectiveness in real-world settings, with the majority of users preferring our approach for its intuitive directional guidance and low inference time. While outdoor navigation yielded higher satisfaction due to fewer obstacles and broader spaces, indoor navigation posed challenges that require further refinement.

Despite the promising results, our study identified several limitations, including challenges in low-light environments, dynamic obstacle handling, and long-distance navigation. Future research should focus on integrating multi-modal sensor inputs, improving adaptability to rapidly changing environments, and refining global path-planning mechanisms to enhance user autonomy. Additionally, addressing accessibility barriers through collaborations with BLV communities will be essential in optimizing the system’s practical adoption.

\bibliographystyle{ACM-Reference-Format}
\bibliography{sample-base}

\section*{Appendix}

\subsection*{A. Application Design Decisions}
\label{lab:design}

This section summarizes the design of the haptic feedback mechanism used in the application. 
A clock-based directional feedback system was adopted for intuitive navigation, where each 
clock position corresponds to a specific vibration pattern. Short pulses represent directions 
from 12:00 to 3:00, while long pulses represent directions from 9:00 to 11:30. The mapping 
between clock time, vibration sequence, and vibration type is shown in 
Table~\ref{tab:vibration-clock-full}.

\begin{table}[h]
\centering
\caption{Vibration Patterns and Types for Clock-Based Directional Feedback}
\label{tab:vibration-clock-full}
\begin{tabular}{|p{1cm}|p{4.3cm}|c|}
\hline
\textbf{Clock Time} & \textbf{Vibration Pattern (ms)} & \textbf{Vibration Type} \\
\hline
12:00 & [0, 200] & Single Short Pulse \\
12:30 & [0, 200] & Single Short Pulse \\
1:00 & [0, 200, 100, 200] & Double Short Pulse \\
1:30 & [0, 200, 100, 200] & Double Short Pulse \\
2:00 & [0, 200, 100, 200, 100, 200] & Triple Short Pulse \\
2:30 & [0, 200, 100, 200, 100, 200] & Triple Short Pulse \\
3:00 & [0, 200, 100, 200, 100, 200, 100, 200] & Quad Short Pulse \\
9:00 & [0, 600, 100, 600, 100, 600] & Triple Long Pulse \\
9:30 & [0, 600, 100, 600, 100, 600] & Triple Long Pulse \\
10:00 & [0, 600, 100, 600] & Double Long Pulse \\
10:30 & [0, 600, 100, 600] & Double Long Pulse \\
11:00 & [0, 600] & Single Long Pulse \\
11:30 & [0, 600] & Single Long Pulse \\
\hline
\end{tabular}
\end{table}

\subsection*{B. User Study}
\label{lab:study}

A usability study was conducted with 15 participants with different blindness profiles, 
mobility aids, and levels of experience. The study evaluated ease of use, response times, 
and preferences for navigation schemes. Tables~\ref{tab:usability_study} and 
\ref{tab:selected_usability_data} summarize participant demographics, interaction 
approaches, and user feedback. 

\begin{table*}[h]
\centering
\caption{Demographic details, mobility aids, approach procedures, ease of use, response times, and method rankings from participant feedback.}
\label{tab:usability_study}
\resizebox{\textwidth}{!}{
\begin{tabular}{l l l l l l l l l l l}
\toprule
\textbf{PID} & \textbf{Blindness Profile} & \textbf{Age} & \textbf{Gender} & \textbf{Education} & \textbf{Location} & \textbf{Mobility Aid} & \textbf{Participant Interaction Approach} & \textbf{Ease of Use} & \textbf{Response Time} & \textbf{Method Ranking} \\
\midrule
1 & Blind by birth & 25 & Male & No Formal Education & Faridpur & White Cane & Talk outside the eye hospital in Faridpur & Average & 2 minutes & 6>3>5>4>2>1 \\
2 & Blind due to accident & 47 & Female & Higher Secondary & Dhaka & White Cane & Request for permission to use the app & Average & 3 minutes & 6>3>4>5>1>2 \\
3 & Due to Glaucoma & 74 & Male & Retired School Teacher & Noakhali & White Cane & Requested a school teacher's feedback & Good & 4 minutes & 6>3>4>5>2>1 \\
4 & Diabetic retinopathy & 54 & Female & Secondary Education & Barishal & Smart Cane & Talked to a stranger at Faridpur Diabetic Hospital & Average & 4 minutes & 6>5>3>4>2>1 \\
5 & Stroke & 69 & Male & Retired Doctor & Khulna & Haptic Feedback Devices & Requested permission to use the app & Good & 2 minutes & 6>5>3>4>1>2 \\
6 & Vitamin A deficiency & 62 & Male & Former NGO Worker & Faridpur & Smart Cane & Collected feedback from a passerby on Newmarket Street & Good & 5 minutes & 5>6>4>3>2>1 \\
7 & Stroke & 71 & Female & Retired Lawyer & Faridpur & White Cane & Collected feedback from a patient at BSMRH Hospital & Bad & 4 minutes & 6>3>4>1>2>5 \\
8 & Macular degeneration & 80 & Male & Retired College Professor & Jessore & GPS-Based Smart Assistants & Asked a neighbor for feedback & Average & 2 minutes & 3>5>6>1>2>4 \\
9 & Blind due to accident & 48 & Male & Vocational Education & Jessore & Guide Dog & Asked a friend for feedback & Good & 6 minutes & 6>3>5>4>2>1 \\
10 & Cataract (Untreated) & 65 & Male & Primary Education & Chattogram & White Cane & Talked to a shopkeeper near his house & Good & 3 minutes & 6>3>4>5>1>2 \\
11 & Diabetic Retinopathy & 52 & Female & Secondary Education & Sylhet & GPS-Based Smart Assistants & Discussed in a community center & Bad & 5 minutes & 5>6>3>4>2>1 \\
12 & Cataracts & 70 & Male & Higher Secondary & Rajshahi & White Cane & Spoke with a friend at the bus station & Good & 4 minutes & 6>5>4>3>2>1 \\
13 & Retinitis Pigmentosa & 35 & Male & Bachelor's Degree & Rangpur & Smart Cane & Collected feedback at a local café & Good & 3 minutes & 6>4>3>5>1>2 \\
14 & Birth Condition & 29 & Female & University Graduate & Mymensingh & Guide Dog & Asked a university friend for feedback & Good & 2 minutes & 6>5>3>4>2>1 \\
15 & Age-related vision loss & 76 & Male & Secondary School & Mymensingh & Haptic Feedback Devices & Tested the app while walking with a family member & Good & 6 minutes & 6>3>4>1>2>5 \\
\bottomrule
\end{tabular}}
\end{table*}

\begin{table*}[h]
\centering
\caption{Participant demographics, navigation scheme preferences, and user feedback on the usability of the app.}
\label{tab:selected_usability_data}
\resizebox{\textwidth}{!}{
\begin{tabular}{l l l l l l p{8cm}}
\toprule
\textbf{PID} & \textbf{Vision Ability} & \textbf{Scene Type} & \textbf{Time to Learn App} & \textbf{Duration of Use} & \textbf{Navigation Scheme Ranking} & \textbf{User Feedback} \\
\midrule
1 & Totally blind & Outdoor & 80 secs & 6 minutes & Left/Forward/Right > Clock Hour Hand > In Degree & This method is well-suited for complex scenarios while maintaining a lower inference time. The 6th method is the most optimal choice for low latency. \\
2 & Light perception only & Indoor & 40 secs & 10 minutes & Clock Hour Hand > Left/Forward/Right > In Degree & I think this method is tailored for outdoor scenarios, ensuring efficient performance in dynamic environments. \\
3 & Tunnel vision & Indoor & 50 secs & 12 minutes & Clock Hour Hand > Left/Forward/Right > In Degree & I feel like the method which took the least amount of time is perfect. The 6th method is my choice for low latency. \\
4 & Blurry vision & Outdoor & 70 secs & 8 minutes & Clock Hour Hand > Left/Forward/Right > In Degree & The clock hour-hand scheme is an effective way to represent direction. The 6th method stands out for its efficiency. \\
5 & Partial vision loss & Indoor & 45 secs & 12 minutes & Clock Hour Hand > In Degree > Left/Forward/Right & The clock hour-hand scheme is more intuitive, but the 6th method balances efficiency and accuracy. \\
6 & Difficulty seeing at night & Outdoor & 30 secs & 10 minutes & In Degree > Clock Hour Hand > Left/Forward/Right & Method 5 has the lowest inference time, ideal for fast applications. Method 6 is more balanced but not the fastest. \\
7 & Partial vision loss & Outdoor & 40 secs & 9 minutes & Clock Hour Hand > Left/Forward/Right > In Degree & Method 6 delivers better results in most cases. I appreciate its way of representing direction. \\
8 & Distorted vision & Indoor & 55 secs & 10 minutes & Clock Hour Hand > In Degree > Left/Forward/Right & I prefer method 3 as it provides better results, and using the clock hour-hand direction is a good choice. \\
9 & Light perception only & Indoor & 50 secs & 14 minutes & Left/Forward/Right > Clock Hour Hand > In Degree & The left/forward/right scheme is easier to understand. Approach 6 is innovative and provides a balanced result. \\
10 & Totally blind & Outdoor & 75 secs & 10 minutes & Left/Forward/Right > Clock Hour Hand > In Degree & The left/forward/right method feels natural, like how someone would guide me in real life. Method 6 worked really well. \\
11 & Blurry vision & Outdoor & 65 secs & 12 minutes & Left/Forward/Right > Clock Hour Hand > In Degree & The left/forward/right scheme makes the most sense to me, but the clock hour-hand system sometimes requires more thought. \\
12 & Partial vision loss & Outdoor & 45 secs & 9 minutes & Clock Hour Hand > Left/Forward/Right > In Degree & The clock hour-hand method was the easiest to follow in real-life situations, providing fast and clear directions. \\
13 & Tunnel Vision & Outdoor & 50 secs & 8 minutes & In Degree > Clock Hour Hand > Left/Forward/Right & I liked method 6 because it felt natural. The degree-based method was interesting, but sometimes required stopping to process. \\
14 & Totally Blind & Indoor & 70 secs & 10 minutes & Left/Forward/Right > Clock Hour Hand > In Degree & I love how the app gives clear instructions. Method 6 was quick, clear, and effortless to understand. \\
15 & Blurry Vision & Indoor & 50 secs & 12 minutes & Left/Forward/Right > Clock Hour Hand > In Degree & The left/forward/right method was the easiest because it felt natural. Method 6 provided clear instructions and confidence. \\
\bottomrule
\end{tabular}}
\end{table*}

\end{document}